
\hsize=6.5in
\vsize=9in
\magnification=1200 
\mathsurround=2pt

\font\Lrg=cmr12 at 22pt 
\font\lrg=cmr9 at 14pt
\font\med=cmr8 at 12pt  
\font\sml=cmti8

           
\font\mib=cmmib10  \textfont8=\mib
\font\smib=cmmib10 at 7pt \scriptfont8=\smib
\mathchardef\balf="080B
\mathchardef\bgam="080D
\def\ib #1{\hbox{\mib #1}}

      \def\p{\ib p} 
\def\r{\ib r}       
\def\v{\ib v}

\font\mil=cmmi6 at 14pt \textfont13=\mil
\mathchardef\psil="0D20  

\font\msa=msam10     \textfont9=\msa
\mathchardef\les="3936   
\mathchardef\ges="393E   
\mathchardef\sqrw="3903  



\def\vep{\varepsilon}
\def\vfi{\varphi}

\def\dlt{\delta}
\def\alf{\alpha}
\def\gam{\gamma}
\def\bPi{{\bf\Pi}}
\def\tta{\theta}


\def\ttxt #1{\qquad\hbox{#1}\qquad}
\def\tld{\tilde}
\def\dag{\dagger}
\def\<{\langle}
\def\>{\rangle}
\def\ovr{\over}

\def\.{\hskip 0.75pt}
\def\'{\hskip -0.75pt}
\def\nbl{\nabla}
\def\tri{\triangle}
\def\tms{\'\times\'}
\def\pnt{\,\raise2pt\hbox{\bf.}\,}
\mathchardef\lb="465B  
\mathchardef\rb="565D  
\def\dcx{d^{\hskip0.7pt 3}\hskip-0.7pt x} 
\def\dll #1{\hbox{\rm l\hskip-1.5pt #1}} 

\def\d{\partial}
\def\od #1#2{{d#1}\over{d#2}}
\def\pd #1#2{{\partial #1}\over{\partial #2}}


\headline={\ifnum\pageno>1{\sml Complex Space-localized Fields \hfil Bodurov} \else\hfil\fi}

\centerline{\Lrg Complex Space\.-localized Fields}
\vskip 9pt
\centerline{\Lrg Interacting with Electromagnetic Fields}
\vskip 25pt

\centerline{\med Theodore Bodurov}
\vskip 25pt

Several families of nonlinear field equations are known to possess space-localized singularity-free solutions which describe fields with finite Hermitian norms. This paper studies the interaction of such fields $\psi$ with given electromagnetic potentials $A_\mu$ entering the field's equations through the covariant derivative forms \ $\d_\mu\psi-gA_\mu\psi$. \ The main result is: the motion of a space-localized field, as a whole, is essentially identical to that of a point charge moving in the same electromagnetic potentials, if this field is a solution either to a nonlinear Schr\"odinger or to a nonlinear Dirac equation, under some very general and physically meaningful assumptions. Using an alternative method, similar result is shown to hold for a very wide class of nonlinear field equations. All of these establish a new connection between nonlinear field theory, on one hand, and classical electrodynamics and quantum mechanics, on the other.
\vskip 25pt

\noindent{\lrg 1.\ Introduction}
\vskip 9pt

Since the well known work of Lorentz and Abraham on the spatially extended electron several types of space-localized fields (under various names) and their possible role in particle/field theory have been studied by many physicists. Among the most prominent are: L.~de~Broglie [1], W.~Heisenberg [2], T.~D.~Lee [3, 4], N.~Rosen [5, 6]. . .

Our interest in the dynamics of space-localized fields, considered as discrete entities, is due primarily to the novel link which it establishes between nonlinear field theories, on one hand, and classical electrodynamics and quantum mechanics, on the other. In this respect the present paper is closely related to an earlier paper [7] by the same author.

Here we study the global dynamics of space-localized complex-valued multi-component fields which are solutions to a large class of nonlinear wave equations. In these equations, which are derivable from Lagrangian densities, the coordinates and the time enter explicitly only via the covariant derivative forms \ $\d\psi/\d x^\mu+i\.U_{\!\mu}(x)\psi$, \ where $U_{\!\mu}(x)$ are known real-valued functions of the time and space coordinates $x^\mu$, \ $\mu=0,1,2,3$. \ Section~3 derives the dynamics of a space-localized field, as a {\it single entity\/}, when it is a solution to a nonlinear Schr\"odinger equation modified with the substitutions
$${\d\ovr\d x^\mu}\ \to\ {\d\ovr\d x^\mu}-igA_\mu\ ,\quad\mu=0,1,2,3\ ,\eqno(1.1)$$ 
where $g$ is a real constant and $A_\mu$ is the 4-potential of a given electromagnetic field. Section 4 presents the same derivations in the case when the localized field is a solution to a nonlinear Dirac equation modified with (1.1). The derivations and the results in these sections are new. They show that the center of the field's region of localization moves as a classical point charge would move in the same electromagnetic field (given by $A_\mu(x)$) provided the latter's variation within the localization region is sufficiently small. Section 5 demonstrates that these results are very general by deriving them (using an alternative method) from three physically natural assumptions but without assuming that the nonlinear field equation is of a particular type.
 
N. Rosen [5] (page 98) appears to be the first who has inferred, without giving any derivation, a similar result by considering the energy of a space-localized field obeying a specific \hbox{\it gauge invariant\/} wave equation. Much latter, Bialynicki-Birula and J. Mycielski [8] investigating a nonlinear Schr\"odinger equation with a {\it logarithmic\/} nonlinear term found that this equation admits closed-form exact localized solutions, which they called {\it gaussons}. In the same paper they have demonstrated that if their Logarithmic NLS equation is modified with the substitution (1.1) then \ {\it ``.\..\..\.in every electromagnetic field, sufficiently small gaussons move like (charged) classical particles.''} 
\vskip 5pt

{\bf Notation and conventions:} \ Here, all complex-valued fields will be denoted with the Greek letters $\psi$, $\vfi$ and $\eta$, all densities (the integrands of functionals) with script capital letters, all 3-vectors with bold letters. Latin indeces will be used for the field's components and for the space coordinates. Greek indeces will be used for space-time components. The complex conjugate of any field $\psi$ will be written as $\psi^*$.
The rest of the symbols will be defined at their first appearance. The summation convention of repeated indexes is assumed as usual. The domain of all space integrals is the entire \hbox{3\.-\.dimensional} Euclidian space ${\rm l\!R}^3$. The units are selected so that \ $c=1$.
\vskip 25pt

\noindent{\lrg 2.\ The assumptions} 
\vskip 9pt

Both for clarity and for preciseness the meaning of the term {\it space-localized\/}, as used in this paper, will be given by the following:
\vskip 5pt

{\bf Definition 2.1.} {\it A scalar, vector or spinor-valued function \ $\psi=(\psi_1\.,\ldots,\psi_s)$ \ whose components \ $\psi_k=\psi_k(x,t)$ \ are functions of the coordinates \ $x=(x^1,x^2,x^3)$ \ and possibly the time $t$ will be called space-localized, or $\psi\!$-field, if its Hermitian norm
$$N=\<\psi,\psi\>=\!\int\limits_{{\rm l\!R}^3}\!\psi_k^*\,\psi_k\,\dcx\ ,
\qquad k=1,\ldots,s\eqno(2.1)$$
is finite and if \ $b<\max_x|\psi_k|<\infty$, \ with \ $b>0$ \ a constant, for all \ $t \ges 0$.}

All functionals which appear in this paper are assumed to be finite when evaluated with a space-localized field.

It is known that wave equations in 3 space-dimensions which do not depend on $x$ explicitly may possess space-localized solutions only if they are {\it nonlinear\/}. {\it Wave packets\/} are not localized fields since they dissipate in time, i.e., for them \ $b=0$. \ We will assume that in a $\psi\!$-field there is only a single {\it region of localization\/}, or that if several such regions are present they are separated by distances which are much larger than their characteristic dimensions. Such an assumption will not diminish the generality of the following derivations, but it will allow us to focus on the key question: {\it How is the motion of the localization region of $\psi$ determined by a given external field, i.e., by the explicit dependence of the $\psi$ wave equation on $x$ and $t$?}

When the $\psi\!$-field equations are obtained from a Lagrangian density ${\cal L}$ and certain assumptions are met it is possible to derive either from the field equations (Sec.\ 3 and 4) or solely from the  Lagrangian density ${\cal L}$ (Sec.\ 5) the differential equations which the coordinates \ $r^i=r^i(t)$ \ of the {\it ``center''\/} of localization satisfy. In essence, the first assumption is: The external potentials enter the nonlinear field equations exactly as the electromagnetic potentials enter the fundamental equations of quantum mechanics.

{\bf Assumption 2.1.} \ {\it The explicit dependence, when present, of the Lagrangian density
$${\cal L}={\cal L}(\psi^*\!,\psi,\d\psi^*\!,\d\psi,t,x)
={\cal L}(\psi^*\!,\psi,\eta^*\!,\eta)\eqno(2.2)$$
on the coordinates $x^i$ and on the time $t=x^0$ is only through the forms \ $\eta=\{\eta_{k\mu}\}$}:  
$$\eta_{k\mu}={\pd{\psi_k}x^\mu}+i\.U_{\!\mu}\.\psi_k\ ,\qquad 
k=1,\ldots,s\ ;\quad \mu=0,1,2,3\eqno(2.3)$$
where \ $U_{\!\mu}=U_{\!\mu}(x,t)$ \ are given real-valued functions, \ $\psi$ may be a spinor field, a vector field, or just a set of coupled scalar fields, \ $\d\psi$ stands for the set of all space and time derivatives \ $\d_\mu\psi_k=\d\psi_k/\d x^\mu$. The behavior of $U_{\!\mu}$ and that of the field components $\psi_k$ under rotations in 3 and 4-space are left undetermined in order to maintain generality. However, the field equations obtained from (2.2) must be Galilei or Lorentz invariant.

The above assumption raises the following question: If a nonlinear wave equation possesses a space-localized solution, will it possess such a solution after it is modified with the substitution (1.1)? This question is answered at the end of this section.

{\bf Assumption 2.2.} \ {\it The functions \ $U_{\!\mu}(x,t)$ \ are arbitrary except that the magnitudes of their variations \ $|\tri U_{\!\mu}|$ \ within a localization region are sufficiently small, so that within such a region $U_{\!\mu}$ may be treated as being constant.}

A criterion for ``when are the variations $|\tri U_{\!\mu}|$ sufficiently small'' is given at the end of Section 5.

{\bf Assumption 2.3.} \ {\it The Lagrangian density \ ${\cal L}(\psi^*\!,\psi,\d\psi^*\!,\d\psi)$ \ obtained from 
(2.2) by setting \ $U_{\!\mu}=0$ \ in (2.3) is gauge type I invariant, that is, the transformation
$$\psi_k'=e^{\.i\.\vep}\,\psi_k\ ,\qquad\psi_k'^*=e^{-\.i\.\vep}\,\psi_k^*\eqno(2.4)$$
with real parameter $\vep$ (which does not depend on $x$ or on $t$) preserves the value of ${\cal L}$.}

Here, the term {\it ``gauge type I transformation''} is used as in Goldstein [9] (page 593) to distinguish from the {\it gauge transformations\/} of the electromagnetic potentials.

Assumption~2.3 is a natural extension of a fundamental property of Schr\"odinger and Dirac equations to the nonlinear field equations, considered in this paper.

When Assumption 2.3 holds it follows from Noether's theorem that there exist a conserved density ${\cal Q}$ and a corresponding flux density \ ${\cal F}=({\cal F}^1\',{\cal F}^2\',{\cal F}^3)$ \ associated with any ${\cal L}$ of the form (2.2) and given by
$${\cal Q}=i\.\bigg(\psi_k^*\,{\pd{\cal L}\eta_{k0}^*}
-\psi_k\,{\pd{\cal L}\eta_{k0}}\bigg)\ ,\qquad
{\cal F}^{\.i}=i\.\bigg(\psi_k^*\,{\pd{\cal L}\eta_{k i}^*}
-\psi_k\,{\pd{\cal L}\eta_{k i}}\bigg)\eqno(2.5)$$
where $\eta_{k0}$, $\eta_{ki}$ are defined by (2.3) and $\psi$ is a solution of the field equation derived from $\cal L$. In view of the critical role which the expressions (2.5) play here, we will sketch an alternative derivation which does not rely on Noether's theorem.

Let \ ${\cal L\.}'={\cal L}(\psi'^*\!,\psi',\eta'^*\!,\eta')$ \ be the Lagrangian obtained from (2.2) by transforming the $\psi\!$-field according to (2.4). Then, the invariance condition is
$${\od{{\cal L\.}'}\vep}\bigg|_{\,\vep\.=\.0}\!
=i\.\psi_k\.{\pd{\cal L}\psi_k}-i\.\psi_k^*\.{\pd{\cal L}\psi_k^*}
+i\.\eta_{k\mu}\.{\pd{\cal L}\eta_{k\mu}}
-i\.\eta_{k\mu}^*\.{\pd{\cal L}\eta_{k\mu}^*}=0\ .\eqno(2.6)$$
Adding to this equation the Euler-Lagrange equation
$${\pd{\cal L}\psi_k}+i\.U_{\!\mu}\.{\pd{\cal L}\eta_{k\mu}}
-{d\ovr dx^\mu}\.{\pd{\cal L}\eta_{k\mu}}=0\eqno(2.7)$$
multiplied by $-\.i\.\psi_k$ and the complex conjugate of (2.7) multiplied by $i\.\psi_k^*$ produces immediately the conservation law
$${\od{\cal Q}t}+\nbl\pnt{\cal F}=0\eqno(2.8)$$
and the expressions (2.5) for which (2.8) holds.

In what follows we will need a variable \ ${\ib r}=(r^1\',r^2\',r^3) \in {\dll R}^3$ \ which gives the {\it position\/} of the localized $\psi\!$-field as a whole. This variable is defined, just as the position of any distribution, with the functionals
$$r^i={1\ovr Q}\int\limits_{{\rm l\!R}^3}\!{\cal Q}\,x^i\,\dcx\ ,\ttxt{with} 
Q=\!\int\limits_{{\rm l\!R}^3}\!{\cal Q}\,\dcx\ ,\quad i=1,2,3\eqno(2.9)$$
where ${\cal Q}$ is given by (2.5) and $Q$ is a constant of the motion according to (2.8).
The values of the functionals $r^i$ are functions of $t$ only and may be regarded as the coordinates of the {\it ``center''\/} of the localization region. No probabilistic interpretation will be imposed on the density ${\cal Q}$, or on any other density, since here such would be entirely unnecessary. The integral in (2.9) for $r^i$ is normalized by dividing it with $Q$ since the $\psi\!$-field, being a solution of a nonlinear equation in general, will not remain a solution if it is normalized. The curve \ $\r=\r(t)$ \ will be called {\it trajectory\/} of the localization region.

If a ``free'' wave equation obtained from the Lagrangian (2.2) with \ $U_{\!\mu}=0$ \ possesses a localized solution, when $U_{\!\mu}$ are arbitrary functions no localized solution will exist, in general. However, if $U_{\!\mu}$ are strictly constant, but otherwise arbitrary, within the space occupied by the $\psi\!$-field then a localized solution will certainly exist. To see this, apply to the Lagrangian (2.2) the transformation \ $\psi'_k=\exp(i\.\tta)\,\psi_k$ \ in which \ $\tta=U_{\!\mu}\,x^\mu$ \ (not a Gauge type I transformation), where all \ $U_{\!\mu}=\rm const$. \ Then, observing that the forms $\eta_{k\mu}$ (2.3) when written in terms of $\psi'_k$ become
$$\eta_{k\mu}={\pd{\psi_k}x^\mu}+i\.U_{\!\mu}\.\psi_k
=e^{-\.i\.\tta}\,{\pd{\psi'_k}x^\mu}$$
we apply Assumption 2.3 (for which the dependence of $\tta$ on $x$ and $t$ is now irrelevant) to obtain the equality
$${\cal L}(\psi^*\!,\psi,\eta^*\!,\eta)
={\cal L}(e^{i\.\tta}\psi'^*\!,\,e^{-\.i\.\tta}\psi',\,e^{i\.\tta}\d\psi'^*\!,\,e^{-\.i\.\tta}\d\psi')
={\cal L}(\psi'^*\!,\psi',\d\psi'^*\!,\d\psi')\ .$$
This shows that if we set \ $U_{\!\mu}=0$ \ in the equations for $\psi$ (but not in $\tta$) we get the equations for $\psi'$. Recognizing that the Hermitian norm of $\psi$ is not affected by the transformation \ $\psi'_k=\exp(i\.\tta)\,\psi_k$ \ the above claim follows.

Now, continuity considerations lead us to conclude that even when $U_{\!\mu}$ are not strictly constant the localized solution will still exist, as long as the  variation's magnitudes \ $|\tri U_{\!\mu}|$ \ within the region of localization remain smaller than a certain critical value $k_1$. This is one reason for having Assumption 2.2. The second reason is that a Lagrangian function for the coordinates $r^i(t)$ can be defined, as we will see in Section 5, only when \ $|\tri U_{\!\mu}|$ \ is smaller than some other limit value \ $k_2$. It may be argued that \ $k_1=k_2$, \ but for our purposes this is not needed.
\vskip 25pt

\noindent{\lrg 3.\ The localized field obeys a nonlinear Schr\"odinger equation} 
\vskip 9pt

Consider the family of nonlinear Schr\"odinger (NLS) equations
$$i\.\hbar\.{\pd{\psi}t}=-\,{\hbar^2\ovr2\.m}\.\nbl^2\psi+\psi\,G(\psi^*\psi)\eqno(3.1)$$
for the scalar-valued field $\psi$. The various constants which appear in Schr\"odinger equation proper have been retained in (3.1), so that a comparison with certain results from QM can be made most conveniently. It is known that there are functions \ $G=G(\psi^*\psi)$ \ with which (3.1) possesses space-localized solutions. For the existence conditions and the proofs see Berestycki and Lions [12].

The Lagrangian density for the NLS equations (3.1) is
$${\cal L}={i\ovr2}\Big(\psi^*{\pd\psi t}-{\pd{\psi^*}t}\.\psi\Big)-{\hbar\ovr2\.m}\.\nbl\psi^*\pnt\nbl\psi
-{1\ovr\hbar}\.{\cal G}(\psi^*\psi)\eqno(3.2)$$
where \ $G(\rho)=d\.{\cal G}(\rho)/d\rho$. In the presence of an electromagnetic field the substitution (1.1) with \ $g=q/\hbar$ \ is inserted in (3.2), according to Assumption 2.1. Then, equation (3.1) becomes
$$i\.\hbar\.{\pd{\psi}t}=\Big(\.{m\ovr2}\,{\bf V}\pnt{\bf V}
+q\.\Phi+G(\psi^*\psi)\'\Big)\psi\ ,\qquad{\bf V}=-\.{1\ovr m}(\.i\.\hbar\nbl+q{\bf A})\eqno(3.3)$$
which will also have localized solutions if Assumption 3.2 is met (see the discussion at the end of Section 2). The first of expressions (2.5) gives \ ${\cal Q}=\psi^*\psi$. Thus, the coordinates $r^i$ of the center of the localization region, according to (2.9), are
$$r^i={1\ovr N}\!\int\limits_{{\rm l\!R}^3}\!\psi^*\psi\,x^i\,\dcx\ ,\qquad 
N=\!\int\limits_{{\rm l\!R}^3}\!\psi^*\psi\,\dcx\ .\eqno(3.4)$$
With the use of (3.3) the velocity \ ${\ib v}=(v^1,v^2,v^3)$ \ of the localization region is 
$$\eqalignno{
{\ib v}={d{\ib r}\ovr dt}&={1\ovr N}\!\int\!\Big(\psi^*{\ib x}\,{\pd\psi t}
+{\pd{\psi^*}t}\.{\ib x}\,\psi\Big)\dcx\cr&={i\ovr\hbar N}\!\int\!\psi^*
\Big(\.{m\ovr2}\big[{\bf V}\pnt{\bf V}\',{\ib x}\big]
+\big[G,{\ib x}\big]\Big)\psi\,\dcx
={1\ovr N}\!\int\!\psi^*\.{\bf V}\psi\,\dcx &(3.5)}$$
since the commutator \ $\big[{\bf V}\pnt{\bf V}\',{\ib x}\big]$ \ gives
 $$\big[{\bf V}\pnt{\bf V}\',{\ib x}\big]
=\big[{\rm V}^j{\rm V}^j\.,{\ib x}\big]
={\rm V}^j\.\big[{\rm V}^j\.,{\ib x}\big]
+\big[{\rm V}^j\.,{\ib x}\big]\.{\rm V}^j={2\.\hbar\ovr i\.m}\.{\bf V}$$
and since the function $G(\psi^*\psi)$ certainly commutes with ${\ib x}$.

According to (3.5) $\bf V$ is the {\it velocity operator\/} for a NLS equation and it is the same as the velocity operator in QM. We use again (3.3) to find the second time derivative of ${\ib r}$
$$\eqalignno
{&{d^2{\ib r}\ovr dt^2}={d\.{\ib v}\ovr dt}
={1\ovr N}\!\int\!\Big(\psi^*\.{\bf V}\.{\pd\psi t}
+{\pd{\psi^*}t}{\bf V}\psi+\psi^*{\pd{\.\bf V}t}\,\psi\Big)\,\dcx &(3.6)\cr
&={1\ovr N}\!\int\!\psi^*\bigg({m\ovr2\.\hbar\.i}\.
\big[{\bf V},{\bf V}\pnt{\bf V}\big]+{q\ovr i\.\hbar}\.\big[{\bf V},\Phi\big]
-{q\ovr m}{\pd{\bf A}t}\'\bigg)\psi\,\dcx
-{i\ovr\hbar N}\!\int\!\psi^*\big[{\bf V},G\big]\.\psi\,\dcx\ .}$$
For any space-localized function $\psi$ the integral containing the nonlinear term is {\it identically zero\/}, as seen from the following
$$\eqalignno{\int\!\psi^*\big[{\bf V},G\big]\.\psi\,\dcx
&={i\.\hbar\over m}\int\!\big(\psi^*G\,\nbl\psi-\psi^*\nbl(G\,\psi)\big)\,\dcx
={i\.\hbar\ovr m}\int\!G\,\nbl(\psi^*\psi)\,\dcx\cr
&={i\.\hbar\ovr m}\int\!{d\.{\cal G}\ovr d(\psi^*\psi)}\nbl(\psi^*\psi)\,\dcx
={i\.\hbar\ovr m}\int\!\nbl\,{\cal G}(\psi^*\psi)\,\dcx=0\,.&(3.7)}$$
One short and one long calculations, both routine in QM, yield for the remaining two commutators:
$$\eqalignno
{\big[{\bf V},\Phi\big]&=-\,{i\ovr m}\nbl\Phi&(3.8)\cr
\big[{\bf V}\.,{\bf V}\pnt{\bf V}\big]
&={\rm V}^j\.\big[{\bf V}\.,\'{\rm V}^j\big]+\big[{\bf V}\.,\'{\rm V}^j\big]\.{\rm V}^j
={i\.q\ovr m^2}\.\big({\bf V}\tms{\bf B}-{\bf B}\tms{\bf V}\big)\ .&(3.9)}$$
When (3.7), (3.8) and (3.9) are inserted into (3.6) and \ ${\bf E}=-\nbl\Phi-\d{\bf A}/\d t$ \ is taken into account the result is
$$m\.{d^2{\ib r}\ovr dt^2}
={q\ovr N}\!\int\!\psi^*\Big({\bf E}+{1\ovr2}\big({\bf V}\tms{\bf B}
-{\bf B}\tms{\bf V}\big)\'\Big)\psi\,\dcx\ .\eqno(3.10)$$
It is most remarkable that the nonlinear term $G(\psi^*\psi)$ of the NLS equation does not enter in the above equation. Hence, we would have gotten exactly the same result had we taken $\psi$ to be a solution of the Schr\"odinger equation proper ($G\equiv0$). However, the derivation of equation (3.10) is valid only if $\psi$ is a space-localized field. But, in quantum mechanics the only space-localized fields are the {\it bound states\/} in some potential well, and so, the ``center'' coordinates ${\ib r}$ (3.4) of any such bound state are fixed by the potential well location. The {\it wave packets\/} of QM are not localized fields, as explained in Section 2. Thus, the question ``what is the motion of a localized field as a whole?'' is not meaningful in quantum mechanics.

For the space-localized solutions of a NLS equation the above question is certainly meaningful. Its answer would follow immediately from (3.10) had we assumed that the variations of the fields $\bf E$, $\bf B$, instead of those of the potentials $\Phi$, $\bf A$, are negligible within the localization region. (Why Assumption 2.2 was chosen to restrict the potentials and not the fields intensities will become clear in Section 5.) The equations \ ${\ib r}={\ib r}(t)$ \ for the motion of the localization center do follow from equation (3.10) and Assumption 2.2 but the reasoning is not entirely straight forwarded: We change the integration variable $x$ to $x'$ according to \ ${\ib x}={\ib x}'+{\ib r}$, \ where ${\ib r}$ is given by (3.4). Then, the first term in (3.10) becomes (with the abbreviations \ $\Phi(\ib x)\equiv\Phi(\ib x,t)$ \ and \ $\bf A(\ib x)\equiv\bf A(\ib x,t)$)
$$\eqalignno{
&\int\!\psi^*\psi\,{\bf E}({\ib x})\,\dcx
=-\!\int\!\psi^*\psi\Big(\nbl\!_x\.\Phi(\ib x)+{\pd{\bf A(\ib x)}t}\Big)\,\dcx\cr
&=-\.\nbl\!_r\!\!\int\!\psi^*\psi\,\Phi({\ib x}'+{\ib r})\,\dcx'
-\!\int\!\psi^*\psi\,{\d\ovr\d t}{\bf A}({\ib x}'+{\ib r})\,\dcx'\cr
&=-\.\nbl\!_r\Phi({\ib r})\!\int\!\psi^*\psi\,\dcx'
-{\d\ovr\d t}\!\int\!\psi^*\psi\.{\bf A}({\ib x}'+{\ib r})\,\dcx'
+\!\int\!{\bf A}({\ib x}'+{\ib r})\,{\pd{(\psi^*\psi)}t}\,\dcx'\cr
&=-\.\nbl\!_r\Phi({\ib r})\!\int\!\psi^*\psi\,\dcx
-{\pd{{\bf A}({\ib r})}t}\!\int\!\psi^*\psi\,\dcx
={\bf E}({\ib r})\!\int\!\psi^*\psi\,\dcx=N{\bf E}(\ib r)&(3.11)}$$
since within the localization region \ $\Phi({\ib x}'+{\ib r})=\Phi({\ib r})$ \ and \ 
${\bf A}({\ib x}'+{\ib r})={\bf A}({\ib r})$ \ according to Assumption 2.2. Similarly, the term containing $\bf B$ in (3.10) becomes 
$${1\ovr 2}\!\int\!\psi^*\big({\bf V}\tms{\bf B}-{\bf B}\tms{\bf V}\big)\psi\,\dcx
=-\.{\bf B}(\ib r)\tms\!\int\!\psi^*\.{\bf V}\psi\.\dcx=N{\ib v}\tms{\bf B}(\ib r)\eqno(3.12)$$
where \ $\ib v=d{\ib r}/dt$, \ given by (3.5), is the velocity of the localization center. Inserting (3.11) and (3.12) into (3.10) produces the desired result
$$m\.{d^2{\ib r}\ovr dt^2}={q\ovr N}\Big({\bf E}\!\int\!\psi^*\psi\,\dcx
-{\bf B}\tms\!\int\!\psi^*\.{\bf V}\psi\,\dcx\Big)
=q\.\big({\bf E}(\ib r)+{\ib v}\tms{\bf B}(\ib r)\big)\eqno(3.13)$$
which is the equation for the localization center coordinates ${\ib r}(t)$. This equation is identical with the equation for the motion of a classical point charge $q$ with mass $m$ in given\break electric field ${\bf E}({\ib r})$ and magnetic field ${\bf B}({\ib r})$ (see Jackson [10]).
\vskip 25 pt

\noindent{\lrg 4.\ The localized field obeys a nonlinear Dirac equation} 
\vskip 9pt

Now let \ $\psi=(\psi_1\.,\ldots,\psi_4)$ \ be a 4-spinor field which satisfies a nonlinear Dirac (NLD) equation of the form
$$i\.\hbar\.\gam\.^\mu\.{\pd{\psi}x^\mu}-(M_1+\gam\.^5M_2)\.\psi=0\eqno(4.1)$$
where \ $M_1=M_1(s_1\.,s_2)$ \ and \ $M_2=M_2(s_1\.,s_2)$ \ are scalar real-valued functions of the scalar and pseudo-scalar quantities
$$s_1=\bar\psi\.\psi \ttxt{and}s_2=\bar\psi\.\gam\.^5\psi
\qquad (\,\bar\psi=\psi^\dag\gam\.^0\,)\ .\eqno(4.2)$$
As explained in Section 2, all wave equations considered in the present paper must be derivable from a Lagrangian density. If one takes for the above two functions:
$$M_1={\pd{{\cal M}(s_1\.,s_2)}s_1}\ttxt{and}M_2
={\pd{{\cal M}(s_1\.,s_2)}s_2}\eqno(4.3)$$
then, the real-valued Lagrangian density from which (4.1) is obtained has the form
$${\cal L}(0)={i\.\hbar\ovr2}\.\Big(\bar\psi\.\gam\.^\mu{\pd{\psi}x^\mu}
-{\pd{\bar\psi}x^\mu}\.\gam\.^\mu\.\psi\Big)-{\cal M}\eqno(4.4)$$
where \ ${\cal M}={\cal M}(s_1\.,s_2)$ \ is some scalar real-valued function of $s_1$ and $s_2$ and nothing else. It is known that, with certain ${\cal M}$, (4.1) possesses space-localized solutions: Cazenave and Vazquez [13], Soler [14]. Additional references may be found in both of this works. 

When external EM potentials \ $\Phi=-A_0$, \ ${\bf A}=(A_1,A_2,A_3)$ \ are present the equation (4.1) is modified with the substitution (1.1), according to Assumption 2.1. Space-localized solutions will exist if the original equation (4.1) possesses such solutions and if Assumption 2.2 is met (by the argument discussed at the end of Section 2). Equation (4.1) thus modified and written as a Hamiltonian evolution equation becomes
$$i\.\hbar\.{\pd\psi t}=-\,\balf\pnt(i\.\hbar\nbl+q{\bf A})\.\psi
+q\.\Phi\.\psi+M\psi=\big(\balf\pnt\bPi+q\.\Phi+M\big)\psi\eqno(4.5)$$
where \ $\bPi=-\.i\.\hbar\nbl-q{\bf A}$, \ $\balf=(\alf^1,\alf^2,\alf^3)$ \ with \ $\alf\.^i=\gam\.^0\.\gam\.^i$, \ and
$$M=\gam\.^0\big(M_1+\gam\.^5M_2\big)\ .\eqno(4.6)$$
The various constants in equation (4.5) are written just as they appear in its linear version so that a comparison with certain results in Dirac's theory can be made most conveniently. We do the substitution (1.1) in the Lagrangian (4.4) and insert the result into the expressions (2.5). Then it follows that the norm $N$, given by (2.1), is a constant of the motion as it is in the linear case.

The position of the localization center \ ${\ib r}=(r^1,r^2,r^3)$ \ is defined again by
$$r^i={1\ovr N}\!\int\limits_{{\rm l\!R}^3}\!\psi_k^*\psi_k\,x^i\,\dcx
={1\ovr N}\!\int\limits_{{\rm l\!R}^3}\!\psi^\dag\psi\,x^i\,\dcx\ .\eqno(4.7)$$
Then, the velocity of the localization region \ ${\ib v}=(v^1,v^2,v^3)$, \ with the use of (4.5), is
$$\eqalignno{v^i={\od{r^i}t}
&={1\ovr N}\!\int\!\Big(\psi^\dag x^i\.{\pd{\psi}t}+{\pd{\psi^\dag}t}\.x^i\,\psi\Big)\.\dcx\cr
&=-\.{1\ovr N}\!\int\!\psi^\dag\big[x^i\.,\balf\pnt\nbl\big]\psi\,\dcx
={1\ovr N}\!\int\!\psi^\dag\alf^i\.\psi\,\dcx &(4.8)}$$
which is the same as the expectation value of the velocity operator $\balf$ in Dirac's theory because the function $M$, as a multiplication operator, obviously commutes with $x^i$.

In the non-relativistic case the {\it ``mechanical'' momentum\/} of a space-localized field is obtained, of course, from the velocity functional (3.5) by multiplying it with the mass $m$. In Dirac's theory one takes for the ``mechanical'' momentum exactly the same functional, except in it $\psi$ is a spinor-valued field. Here we do the same, i.e., we take
$$\p=-\.{1\ovr N}\!\int\!\psi^\dag(\.i\.\hbar\nbl+q{\bf A})\.\psi\,\dcx
={1\ovr N}\!\int\!\psi^\dag{\bf\Pi}\.\psi\,\dcx\eqno(4.9)$$
to be the ``mechanical'' momentum of the localized field, where \ ${\bf\Pi}=-\.i\.\hbar\nbl-q{\bf A}$ \ is the corresponding operator in both the relativistic and non-relativistic QM, Messiah [15]. (The term {\it``mechanical'' momentum\/} is used as in references 9 and 15.) Now, instead of calculating \ $d{\ib v}/dt$ \ which was done in Section 3, here we calculate \ $d\p/dt$ as follows
$$\eqalignno{{d\p\ovr dt}
&={1\ovr N}\!\int\!\bigg(\psi^\dag\.\bPi\.{\pd{\psi}t}+
{\pd{\psi^\dag}t}\bPi\.\psi+\psi^\dag\.{\pd{\bPi}t}\.\psi\'\bigg)\.\dcx\cr
&={i\ovr\hbar\.N}\!\int\!\psi^\dag\Big(\big[\balf\pnt\bPi\.,\bPi\big]
+q\.\big[\Phi\.,\bPi\big]+\big[M,\bPi\big]
+i\.q\.\hbar\.{\pd{\bf A}t}\Big)\psi\,\dcx\ .&(4.10)}$$
For the first and the second commutators we give only the final results since they are well known in QM, i.e.,
$$\eqalignno{\big[\balf\pnt\bPi\.,\bPi\big]&=\alf^i\.\big[\Pi_i\.,\bPi\big]
=-\.i\.q\.\hbar\,\balf\tms(\nbl\tms{\bf A})
=-\.i\.q\.\hbar\,\balf\tms{\bf B}&(4.11)\cr
\big[\Phi\.,\bPi\big]&=i\.\hbar\.\nbl\Phi\ .&(4.12)}$$
Although the calculation involving the nonlinear term \ $\psi^\dag\big[M,\bPi\big]\psi$ \ is not complicated it is presented below in some detail because it is critical for the present derivation. Indeed, the following will show that the integral of this term over all space is {\it identically\/} zero: 

Substituting the expression for $M$ from (4.6) and taking in consideration (4.3)
$$\eqalignno{&\!\int\!\psi^\dag\big[M,\bPi\big]\psi\,\dcx
=-\.i\.\hbar\!\int\!\big(\psi^\dag M\nbl\psi
+(\nbl\psi^\dag)M\psi\big)\,\dcx\cr
&={\hbar\ovr i}\int\!\Big(\big(\.(\nbl\psi^\dag)\gam\.^0\psi
+\psi^\dag\gam\.^0\.\nbl\psi\big)M_1
+\big(\.(\nbl\psi^\dag)\gam\.^0\gam\.^5\psi
+\psi^\dag\gam\.^0\gam\.^5\.\nbl\psi\big)M_2\Big)\,\dcx\cr
&={\hbar\ovr i}\int\!\Big(M_1\nbl(\psi^\dag\gam\.^0\psi)
+M_2\nbl(\psi^\dag\gam\.^0\gam\.^5\psi)\Big)\,\dcx
={\hbar\ovr i}\int\!\bigg({\pd{\cal M}s_1}\nbl s_1
+{\pd{\cal M}s_2}\nbl s_2\bigg)\dcx\cr
&={\hbar\ovr i}\int\!\nbl{\cal M}(s_1\.,s_2)\,\dcx=0&(4.13)}$$
one arrives at the desired result, where as before \ $s_1=\psi^\dag\gam\.^0\psi$ \ and \ $s_2=\psi^\dag\gam\.^0\gam\.^5\psi$. Inserting (4.11), (4.12) and (4.13) into (4.10) and noticing that \ ${\bf E}=-\nbl\Phi-\d{\bf A}/\d t$ \ we finds that
$${d\p\ovr dt}={q\ovr N}\!\int\!\psi^\dag\big({\bf E}
+\balf\tms{\bf B}\big)\psi\,\dcx\eqno(4.14)$$
which is valid for the nonlinear Dirac equations (4.5), as well as for the Dirac equation proper. 
This is so because the integral (4.13), involving the nonlinearity $M$, is zero.

Assumption 2.2 was not needed for the derivation of (4.14). Now, if we assume that the variations of $\Phi$ and $\bf A$ are sufficiently small within the localization region, i.e., if Assumption 2.2 holds, then the electromagnetic fields \hbox{$\bf E$, $\bf B$} may be treated as being constant in that region, as it was shown in Section 3. Hence, (4.14) becomes
$${d\p\ovr dt}={q\ovr N}\Big({\bf E}(\ib r)\!\int\!\psi^\dag\psi\,\dcx
-{\bf B}(\ib r)\tms\!\int\!\psi^\dag\balf\,\psi\,\dcx\Big)
=q\.\big({\bf E}(\ib r)+{\ib v}\tms{\bf B}(\ib r)\big)\eqno(4.15)$$
where \ ${\ib v}=d{\ib r}/dt$, \ given by (4.8), is the velocity of the localization region ``center''. Equation (4.15) describes the motion of the localization region \ $\ib r=\ib r(t)$ \ in complete agreement with the result (3.13) of Section 3.

Again, one recognizes this to be precisely the equation of motion for a point charge $q$ but this time in the relativistic case (Goldstein [9] Sec.\ 7.8, Jackson [10] Sec.\ 12.1).
\vskip 25 pt

\noindent{\lrg 5.\ The Lagrangian function for the motion of a localized field} 
\vskip 9pt

The results in the previous two sections are valid for entire families of NLS and NLD equations which possess space-localized solutions. Thus, the question arises: Is it possible to derive similar results without assuming a particular type for the field equations? This section will show that it is possible, but the method has to be different since one cannot calculate explicitely the time-derivatives of the functionals $\ib r$ nor of $\ib v$ without knowing the field equations. Instead, we will derive from the assumptions in Section~2 the general form of the Lagrangian function for the motion of a space-localized field, i.e., for the coordinates \ $\ib r=(r_1,r_2,r_3)$ \ of the localization center. Then, several important conclusions will be drawn from the form of this Lagrangian function.
  
When Assumption 2.2 holds, i.e. the variations \ $|\tri U_{\!\mu}|$ \ of the given potentials $U_\mu$ are sufficiently small within the localization region, the value of the space integral of the Lagrangian density (2.2) may be approximated by replacing in the latter the actual potentials $U_{\!\mu}(x,t)$ with their values $U_{\!\mu}(r,t)$ at the localization ``center'' $\ib r$. This is so because \ ${\cal L}\to0$ \ outside the localization region. Then, at a fixed time $t$ and for a fixed space-localized $\psi(x,t)$ the Lagrangian functional
$$L\big(U(r,t)\big)=\!\int\limits_{{\rm l\!R}^3}\!
{\cal L}\big(\psi^*\!,\psi,\d\psi^*-iU(r,t)\psi^*,\d\psi+iU(r,t)\psi\big)\,\dcx\eqno(5.1)$$
becomes a function of $U_{\!\mu}$. Alternatively, one can regard (5.1) as a functional of $\psi$ and $\psi^*$ in which $U_{\!\mu}$ enter as {\it parameters\/}. At the end of this section it is shown that the error in this approximation can be made arbitrarily small by choosing potentials whose variations within the localization region are sufficiently small, i.e., \ $|\tri U_0|+|v^i|\'\cdot\'|\tri U_{\'i}|\ll|L/Q|$.

The function $L(U)$ defined by (5.1) is different, of course, for different $\psi$. Our goal is to derive a relationship between $L$ and $U_{\!\mu}$ which remains valid during the entire motion of a particular $\psi$ as long as $\psi$ remains space-localized. To do this we differentiate (5.1) with respect to $U_{\!\mu}$. Because of Assumption 2.2 \ it is permissible to differentiate under the integral sign which produces
$${\pd{L}U_{\!\mu}}=\!\int\!{\pd{\cal L}U_{\!\mu}}\,\dcx
=i\!\int\!\bigg(\!\psi_k\.{\pd{\cal L}\eta_{k\mu}}
-\psi^*_k\.{\pd{\cal L}\eta_{k\mu}^*}\bigg)\,\dcx\ .\eqno(5.2)$$
This is yet another functional of $\psi$ in which $U_{\!\mu}$ enter as parameters. However, {\it if $\psi$ is a space-localized solution of the field equations (2.7), derived from the same Lagrangian density \ ${\cal L}(\psi^*\!,\psi,\eta^*\!,\eta)$, \ and $U_{\!\mu}$ are the values of the potentials at the region of localization, then, the integrands in (5.2) are precisely the conserved density $-\.{\cal Q}$ and the three components of the flux density $-\.{\cal F\.}^i$ as given by (2.5)}. 

Thus, when (5.2) is evaluated on a solution for \ $\mu=0$ \ we have that
$${\pd{L}U_{\!0}}=-\!\int\!{\cal Q}\,\dcx=-\,Q\eqno(5.3)$$
holds during (and only during) the motion, where the quantity $Q$ is a {\it constant\/} during the same motion according to (2.8). Of course, the value of $Q$ is determined by the {\it initial conditions\/} of the particular solution and so $Q$ will be different for different solutions $\psi$.

Again, when (5.2) is evaluated on a solution but now for \ $\mu=i=1,2,3$ \ we have
$${\pd{L}U_{\!i}}=-\!\int\!{\cal F}^{\.i}\.\dcx\ .\eqno(5.4)$$
Next, the velocity \ ${\ib v}=(v^1\',v^2\',v^3)$ \ of the localization region is expressed in terms of $Q$ and ${\cal F}^{\.i}$ as follows: We differentiate the position functionals $r^i$ in (2.9) with respect to $t$
$$v^i={dr^i\ovr dt}={1\ovr Q}\int{d{\cal Q}\ovr dt}\,x^i\,\dcx
=-\,{1\ovr Q}\int{\pd{{\cal F}^j}x^{\.j}}\,x^i\.\dcx
={1\ovr Q}\int\!{\cal F}^{\.i}\,\dcx\eqno(5.5)$$
then we use equation (2.8) and integrate by parts. The last result together with (5.4) yields a relation which holds for the entire motion
$${\pd{L}U_{\! i}}=-\,Q\,v^i\ ,\qquad i=1,2,3\ .\eqno(5.6)$$

Let us now consider the motion of the localization region of a solution $\psi=\psi(x,t)$, assumed to be known, of some nonlinear field equation. We can find the exact {\it trajectory\/} of the localization region \ $\r=\r(t)$ \ by inserting this solution into the functionals (2.9). At any point $\r(t)$ on this trajectory the quantities \ $U_0=U_0(r,t)$ \ and \ $U_i=U_i(r,t)$ \ which appear in (5.3) and (5.6) are well defined and known, since $U_{\!\mu}(x,t)$ are given functions. Furthermore, the velocity \ ${\ib v}(t)=d{\ib r}(t)/dt$ \ is well defined and known, since ${\ib r}(t)$ is known. We want to find a function \ $L=L(U,v)$ \ of the variables \ $U_0\,,\ U_i\,,\ v^k\ (i,k=1,2,3)$ \ such that the equations (5.3) and (5.6) are satisfied during the motion, i.e. at each point of the trajectory \ $\r=\r(t)$. The {\it unique\/} solution of this system of 4 equations, obtained by four consecutive (trivial) integrations with respect to $U_0$, $U_1$, $U_2$ and $U_3$, is
$$L=-\,Q\.\big(U_0+U_{\'i}\,v^i\big)+L(0,v)\eqno(5.7)$$
where $L(0,v)$ depends on $v^i$ but not on $U_{\!\mu}$ and may be regarded as the integration ``constant''. One can verify the expression (5.7) by inserting it in (5.3) and (5.6).

It remains to convince ourselves that $L$ given by (5.7) is the Lagrangian function for the trajectory \ $\r=\r(t)$. \ This is done by restating (5.7) as
$$\int_{{\rm l\!R}^3}{\cal L}(\psi^*\!,\psi,\eta^*\!,\eta)\,\dcx
=L(0,v)-Q\.(U_0+U_{\'i}\,v^i)\eqno(5.8)$$ 
and integrating both sides with respect to time from $t_1$ to $t_2$. Thus, within the error of the approximation the equation
$$\int_{t_1}^{t_2}\!\!\int_{{\rm l\!R}^3}{\cal L}(\psi^*\!,\psi,\eta^*\!,\eta)\,\dcx\,dt
=\int_{t_1}^{t_2}\!\!\big(L(0,v)-Q\.(U_0+U_{\'i}\,v^i)\big)\,dt\eqno(5.9)$$
holds on space-localized solutions $\psi(x,t)$. Moreover, this equation holds also for all space-localized functions $\tilde\psi(x,t)$ which satisfy the conservation law (2.8) and Assumption 2.2, because for such functions (5.8) holds as already shown. Pertinent remarks:

\noindent(a) $\tld\psi$ will be called the {\it varied\/} field if it is sufficiently close in norm to a solution $\psi$.

\noindent(b) $\tld\psi$ is restricted to be a space-localized field since otherwise the Lagrangian functional \ $\int\!{\cal L}\,\dcx$ \ when evaluated with $\tld\psi$ will be divergent, in general, and $L$ will be meaningless.

\noindent(c) The varied field $\tld\psi$ must be such that it represents a particle whose trajectory is varied, but not its properties like mass, charge, etc. Hence, the variations are restricted so that $\tld\psi$ satisfies the conservation law (2.8) and Assumption 2.2. Consequently, $Q$ in (2.9) remains a constant of the motion (which is not varied).

\noindent(d) The functions $U_0(x,t)$ and $U_i(x,t)$ are not varied --- they are given. 

According to the calculus of variations, $\tld\psi(x,t)$ is varied so that its values $\tld\psi(x,t_1)$ at $t_1$ and $\tld\psi(x,t_2)$ at $t_2$ remain fixed, i.e. \ $\tld\psi(x,t_1)=\psi(x,t_1)$ \ and \ $\tld\psi(x,t_2)=\psi(x,t_2)$. Inserting the varied $\tld\psi$ into the expressions  (2.5) \ and the result \ --- \ into  (2.9) \ one finds the varied trajectory \ $\tld{\r}=\tld{\r}(t)$. Thus, by varying the localized field $\tld\psi$ we vary the trajectory. Moreover, since the functions $\tld\psi(x,t_1)$ and $\tld\psi(x,t_2)$ are fixed it follows from (2.9) that the initial point $\tld r^{\.i}(t_1)$ and the final point $\tld r^{\.i}(t_2)$ of the varied trajectory are also fixed, i.e. $\tld r^{\.i}(t_1)=r^i(t_1)$ \ and \ $\tld r^{\.i}(t_2)=r^i(t_2)$, \ just as required by the calculus of variations. 

Having $\tld r^{\.i}(t)$ one can calculate the localization center velocity $\tld v^i(t)$ and the values of $U_0(\tld r,t)$, \ $U_i(\tld r,t)$ \ on the varied trajectory. Finally, one can calculate the value of $L(0,\tld v)$ on the same trajectory by setting $U_0=0$ \ and \ $U_i=0$ \ in (5.1), i.e. from
$$L(0,\tld v)=\!\int\!{\cal L}(\tld\psi^*\!,\tld\psi,\d\tld\psi^*\!,\d\tld\psi)\,\dcx\ .\eqno(5.10)$$
Putting all the above together we see that the expression 
$$L(0,\tld v)-Q\.\big(U_0(\tld r,t)+\tld v^i\.U_{\'i}(\tld r,t)\big)\eqno(5.11)$$
has a well defined value at each point of the varied trajectory. Consequently, the value of the time integral of (5.11) from $t_1$ to $t_2$ is well defined and equation (5.9) holds with any varied field $\tld\psi$ under the above stated restrictions.

Now, consider a $\psi$ for which the left-side integral in (5.9) is at a minimum (maximum). According to the calculus of variations this $\psi$ is a solution of the corresponding Euler-Lagrange equations. Then, as long as equation (5.9) holds, the right-side integral is also at a minimum (maximum) which is achieved for the ``true'' trajectory ${\r}(t)$. This shows that the right-side integral in (5.9) is the {\it action integral\/} for the trajectory ${\r}(t)$ of the localization center and that $L$ given by (5.7) is its Lagrangian function. End of proof.

If we set \ $U_\mu=gA_\mu$ \ in the Lagrangian function (5.7) we see that the first term in it
$$-\,gQ(A_0+v^iA_i)\eqno(5.12)$$
is the part which accounts for the interaction between the space-localized field , as a whole, and the electromagnetic field $A_\mu$. This part is identical with the interaction term in the Lagrangian function for a point charge \ $q=gQ$, \ both in the classical and in the relativistic cases. For the Lagrangian of a point charge see, for example, Goldstein [9] (Section 1.5) or Jackson [10] (Section 12.1). The second part in (5.7), clearly, stands for the Lagrangian function for a ``free'' localized field, as a whole, and it is a function of $\ib v$ only. To find it set \ $U_{\!\mu}=0$ \ in the field equation (now assumed to be known) and obtain the corresponding stationary localized solution. Apply to it a Galilei or Lorentz transformation to obtain a solution whose region of localization is moving with some constant velocity $\v$. Evaluate the Lagrangian functional (5.1) with this transformed solution and with \ $U_{\!\mu}=0$. \ Since $v^i$ are the transformation parameters, the result will be a function of $\v$, as claimed, and nothing else. 
We summarize the first principal result which follows from (5.7):

{\it If Assumptions 2.1, 2.2 and 2.3 are met the interaction term in the Lagrangian\break function (5.7) for the motion of a space-localized $\psi\!$-field, as a whole, is identical to the\break interaction term in the Lagrangian function for a point charge proportional to $Q$ moving in an electromagnetic field proportional to $U_{\!\mu}$, regardless of the type of wave equation which $\psi$ satisfies.}

For the second result which follows from (5.7) we observe that the above statement is valid, in particular, when $\psi$ satisfies a Hamiltonian evolution equation of the form
$$i\,{\pd{\psi_k}t}={\dlt H\ovr\dlt\psi_k^*}\ttxt{with}
H=\!\int\!{\cal H}(\psi^*\!,\psi,\eta^*\!,\eta)\,\dcx\eqno(5.13)$$
where ${\cal H}$ is the Hamiltonian density, $\eta_{k\mu}$ are given by (2.3) and \ ${\dlt H/\dlt\psi_k^*}$ \ is the \hbox{\it variational derivative\/} of the Hamiltonian functional $H$ with respect to $\psi_k^*$. For details see Bodurov [7]. Indeed, it is easy to see that (5.13) is the Euler-Lagrange equation derived from the Lagrangian density (which satisfies Assumptions 2.1, 2.2 and 2.3 if $\cal H$ does):
$${\cal L}={i\ovr2}\,\Big(\psi_k^*\,{\pd{\psi_k}t}
-{\pd{\psi_k^*}t}\,\psi_k\Big)-{\cal H}\ .\eqno(5.14)$$
When $\cal L$ (5.14) is inserted into (2.5) it yields 
$${\cal Q}=\psi_k^*\psi_k \ttxt{and} Q=\<\psi,\psi\>=N\ .\eqno(5.15)$$

The form (5.13) contains, as special cases, Schr\"odinger and Dirac equations together with their nonlinear versions. In these equations the constant $g$ has the value \ $g=q/\hbar$. \ On the other hand, the interaction term (5.12) and expressions (5.15) tell us that the electric charge is \ $q=gN$. \ Hence, we have \ $q/\hbar=g=q/N$ \ from which follows that $\hbar=N$. The dimensionality of $N$ ($\psi$ is not normalized) is certainly right, i.e., [{\it action\/}] as seen from (5.13) by noticing that the dimensionality of $\cal H$ is \ $[{\cal H}]=[energy]/[volume]$
$$[N]=[|\psi|^2]\,[volume]=[{\cal H}]\,[time]\,[volume]=[energy]\,[time]=[action]\ .$$

Thus, our second main result is:

{\it If a space-localized $\psi\!$-field is a solution of a nonlinear complex Hamiltonian evolution equation with the form (5.13) whose Hamiltonian functional is gauge type I invariant, then in certain relationships the Hermitian norm $\<\psi,\psi\>$ appears in the place in which Planck's constant $\hbar$ appears in the corresponding relationships of quantum mechanics.}

Two arguments independent from the one above and pertaining to different relationships but leading to the same conclusion above, were presented in a paper by the same author (Bodurov [11], Sections 3 and 4). A year latter, one of these was discussed with further details in Bodurov [7] (Section 5).

The {\it criterion\/} for the validity of Assumption 2.2, i.e., for the maximum rate of change of $U$ at which equation (5.7) still holds is derived as follows. Let $a$ be an estimate of the {\it radius of localization\/} given by the functional
$$a^2={1\ovr Q}\sum_{i=1}^3\,\int\limits_{\,\,{\rm l\!R}^3}\!
{\cal Q}\.(x^i-r^i)^2\,\dcx\eqno(5.16)$$
where $r^i$ are the coordinates of the ``center'' of localization given by (2.9). Denote by $U_{\max}$ and $U_{\min}$ the values of $U$, within a sphere with a radius $a$ and a center ${\ib r}$, for which the magnitude \ $\big|L(U_{\max})-L(U_{\min})\big|$ \ is maximal. $L(U_{\max})$ and $L(U_{\min})$ are the values of the Lagrangian functional (5.1) when it is evaluated with the constant \ $U=U_{\max}$ \ and \ $U=U_{\min}$. \ Then, the sought criterion for \ $\tri U_{\!\mu}=(U_{\!\mu})_{\max}-(U_{\!\mu})_{\min}$ \ is
$$\big|L(U_{\max})-L(U_{\min})\big|\ll\big|L(U_{\min})\big|\ .\eqno(5.17)$$
(We assume, without loss of physical generality, that $U_{\!\mu}$ are monotonic functions of the coordinates within the localization region.) With the use of (5.3) and (5.6) we can write
$$L(U_{\max})-L(U_{\min}) \approx {\pd{L(U)}U_{\!\mu}}
\bigg|_{\,U_{\min}}\!\!\tri U_{\!\mu}=-\,Q\.\big(\tri U_0+v^i\.\tri U_{\'i}\big)\ .$$  
The last equation together with (5.17) and (5.12) produces \ 
$|\tri U_0|+|v^i|\'\cdot\'|\tri U_{\'i}|\ll|{L(0,v)/Q}|$, \ 
where \ $|\tri U_0|$ \ and \ $|\tri U_{\'i}|$ \ are the variation's magnitudes of the respective potentials over the localization region, $v^i$ are its velocity components and $L(0,v)$ is the value of the Lagrangian for the ``free'' $\psi\!$-field. The above inequality is equivalent to
$$|\tri U_0|\ll\bigg|{L(0,v)\ovr Q}\bigg|\ttxt{and}
|v^i|\'\cdot\'|\tri U_{\'i}|\ll\bigg|{L(0,v)\ovr Q}\bigg|\ ,\quad i=1,2,3\eqno(5.18)$$
that is, the interaction terms in the Lagrangian (5.7) must be much smaller than the ``free'' $\psi\!$-field Lagrangian $L(0,v)$ --- a physically meaningful conclusion.
\vskip 25 pt

\noindent{\lrg 6.\ Concluding discussion} 
\vskip 9pt

It was demonstrated in this paper that if the interaction of a nonlinear space-localized $\psi\!$-field with external electromagnetic potentials is defined according to the substitution (1.1) and Assumptions 2.1, 2.2 and 2.3 are met then the motion of the \hbox{$\psi\!$-field}, as a whole, is identical to that of a classical point charge. Moreover, this identity is largely independent of the particular $\psi\!$-field equation, as long as the latter is derivable from a gauge I invariant Lagrangian. 

The results (3.10) and (4.14) may be regarded as generalizations of Ehrenfest theorem showing that this theorem is valid not only for the linear equations of quantum mechanics but also for the families of NLS and NLD nonlinear field equations which possess space-localized solutions.

One may also say that according to the results (3.13), (4.15) and (5.12) an external electromagnetic field given by a 4-potential, under the Assumptions 2.1, 2.2 and 2.3, acts with the {\it\/ Lorentz ``force''\/} \ $g\.Q\.\big({\bf E}+{\ib v}\tms{\bf B}\big)$ \ on a space-localized $\psi\!$-field, whose equivalent electric charge is \ $q=g\.Q$.

As already discussed, the global description of the time evolution of a $\psi\!$-field by the trajectory of its center of localization \ $\r=\r(t)$ \ is valid when the variations $|\tri U_{\!\mu}|$ within the localization region are sufficiently small. It is worth noticing that this correlates very well with a number of experimental evidences that the concept of elementary particle, as an individual entity, is meaningful only when the intensities of the external fields acting on this particle are within certain bounds.

We would like to look on the results summarized above as new evidence that it is not only possible but it is also very desirable to represent elementary charges with localized $\psi\!$-fields. For, these results point to a unique opportunity to reconcile the {\it discrete aspect\/} with the {\it wave aspect\/} of elementary particles/charges.
\vskip 25pt

\noindent{\lrg References}
\vskip 9pt

\item{[1]\ }L. de Broglie, {\sl Nonlinear Wave Mechanics, Causal Interpretation\/}, Elsevier, 1960.
\item{[2]\ } W. Heisenberg, {\sl Quantum Theory of Fields and Elementary Particles\/},

Reviews of Modern Physics, 29 (1957), No. 3
\item{[3]\ } T. D. Lee, {\sl Particle Physics and Field Theory\/}, Harwood Academic, 1981. 
\item{[4]\ } R. Friedberg, T. D. Lee and A. Sirlin, {\sl Class of Scalar-field Soliton Solutions in Three Space Dimensions,} 
Physical Review D, 13 (1976), No. 10, 2739-2761.
\item{[5]\ } N. \'Rosen, {\sl A Field Theory of Elementary Particles\/},

Physical Review, Vol. \'55 (1939), 94-101.
\item{[6]\ } F. Cooperstock and N. Rosen, {\sl A Nonlinear Gauge-invariant Field Theory of Leptons,}

International Journal of Theoretical Physics, 28 (1989), 423-440.
\item{[7]\ } T. \'Bodurov, {\sl Complex Hamiltonian Evolution Equations and Field Theory\/},

Journal Math. \'Physics Vol. \'39(1998), No. \'11, 5700-5715.
\item{[8]\ } I. \'Bialynicki-Birula and J. \'Mycielski, {\sl Nonlinear Wave Mechanics\/},

Annals of Physics, Vol. \'100 (1976), 62-93.
\item{[9]\ } H. \'Goldstein, {\sl Classical Mechanics\/}, 2nd ed., Addison-Wesley, 1981.
\item{[10]\ } J. \'Jackson, {\sl Classical Electrodynamics\/}, 2nd ed., Wiley, 1975.
\item{[11]\ } T. \'Bodurov, {\sl De Broglie-type Relations from Nonlinear Evolution Equations\/,} 

International Journal of Theoretical Physics, 36 (1997), No. \'8, 1771-1785.

\item{[12]\ } H. \'Berestycki and P. \'Lions, {\sl Nonlinear Scalar Field Equations:

I. Existence of Ground State, \ II. Existence of Infinitely Many Solutions,}

Archive for Rational Mechanics and Analysis, 82 (1983), No. \'4, 313-375.
\item{[13]\ } T. \'Cazenave and L. \'Vazquez, {\sl Existence of Localized Solutions for a Classical Nonlinear Dirac Field\/,} Comm. \'Math. \'Phys., 105 (1986), 35-47.
\item{[14]\ } M. \'Soler, {\sl Classical, Stable, Nonlinear Spinor Field with Positive Rest Energy\/}, 

Physical Review D1 (1970), No. \'10, 2766-2769.
\item{[15]\ } A. \'Messiah, {\sl Quantum Mechanics\/}, Vols. \'\'1, 2, North-Holland, 1962.
\item{[16]\ } U. Enz, {\sl The Sine-Gordon Breather as a Moving Oscillator in the Sense of de Broglie\/,}

Physica 17D (1985) 116-119.
\item{[17]\ } E. Schr\"odinger, {\sl Briefe zur Wellenmechanik\/}, edited by K. Przibram, 

Wien, Springer-Verlag, 1963.

(English translation: \ E. Schr\"odinger, {\sl Letters on Quantum Mechanics\/}, 

edited by K. Przibram, Philosophical Library, 1967, 1986.)
\vfill\eject
\bye